# Pulsed Mid-infrared Radiation from Spectral Broadening in Laser Wakefield Simulations

W. Zhu, J. P. Palastro, and T. M. Antonsen

IREAP, University of Maryland, College Park 20740

Spectral red-shifting of high power laser pulses propagating through underdense plasma can be a source of ultrashort mid-infrared (MIR) radiation. During propagation, a high power laser pulse drives large amplitude plasma waves, depleting the pulse energy. At the same time, the large amplitude plasma wave provides a dynamic dielectric response that leads to spectral shifting. The loss of laser pulse energy and the approximate conservation of laser pulse action imply that spectral red-shifts accompany the depletion. In this paper, we investigate, through simulation, the parametric dependence of MIR generation on pulse energy, initial pulse duration, and plasma density.

# I. Introduction

A short (<100 fs), intense (>$10^{17}$ W/cm$^2$), optical laser pulse propagating through underdense (n~$10^{18}$/cm$^3$) plasma ponderomotively excites plasma waves. Plasma electrons are pushed forward at the pulse front, repelled laterally in the body of the pulse, and converge on axis one plasma period later. As a result, the electron density increases at the front, decreases in the middle and can increase again at the back of the pulse. The density variations of the plasma wave provide a dynamic dielectric response that modifies the pulse spectrum[1-4]: the spectrum blue-shifts where the electron density is rising in time and red-shifts where the electron density is falling in time. For pulse durations commensurate with the plasma period, red-shifting dominates the spectral evolution and occurs where the body of the pulse sits in a region of falling electron density. The red-shifting can lead to a broadband spectrum extending well into the mid-infrared (MIR), wavelengths ranging from 3 to 7 microns, and whose properties are determined by those of the initial pulse and the plasma through which it propagates[5]. Because of its spectral proximity to the natural frequencies associated with molecular vibration, a tunable MIR source could be used to probe fundamental physical processes in liquids and materials or induce time-dependent structural changes in biological assemblies[5,6]. To this end, we investigate the parametric dependence of the conversion from optical to MIR energy in underdense plasma.

The extent of spectral shifting depends on the temporal gradient in the electron density and the distance over which the gradient is sustained[1-5]. The conversion to MIR thus relies critically on increasing and maintaining the ponderomotive force[7-9]. At low powers, preformed plasma channels can provide radial confinement and maintain the ponderomotive force. At higher powers, the radial expulsion of electrons initially focuses the pulse and can result in a transient self-guiding structure: relativistic self-focusing and guiding[10-14]. Additionally, the local reduction in group velocity accompanying the red-shifting compresses the pulse[15-18]. Both nonlinear localization effects can enhance the ponderomotive force and consequently the electron density gradient. If the ponderomotive force is sufficient, complete cavitation of the electron density can occur and is often referred to as the 'bubble regime'[19-21].

Figure 1 illustrates the co-localization of the pulse and plasma wave. In the simulation a pulse with an initial wavelength of 800 nm, energy of 0.5 J, and duration of 30 fs is injected into a plasma of density $4.3 \times 10^{18}$ cm$^{-3}$. The top row display the laser pulse intensity and electron density as a function of radius $r$ and moving frame coodinate $\xi = ct - z$, where $c$ is the speed of light, initially and after 4.9 mm of propagation on the left and right respectively. At 4.9 mm the pulse has undergone significant transverse and temporal compression. The body of the pulse sits in a region where the electron density is falling in time and the region trailing the pulse is completely devoid of electrons. On the bottom row the corresponding Wigner distributions, representations of the local pulse spectrum, are plotted as a function of normalized wavenumber $k/k_0$ and $\xi$. The negative gradient in electron density along the pulse has resulted in the average pulse wavenumber dropping to nearly half its initial value, and the minimum wavenumber extending into the MIR.

This remainder of this paper is organized as follows. Section II describes the ponderomotive guiding center, quasi-static plasma response model and the tenuous plasma, full wave equation (FWE) used in the simulations. In Sec. III, the role of wave action conservation and pulse energy depletion in spectral shifting is discussed. Simulation results for pre-formed, fully ionized plasmas are presented in Sec. IV. In this section, we examine MIR generation for parameters relevant to the University of Maryland laser system, and examine the parametric dependence of MIR energy on laser power, pulse length, and plasma density. Section V concludes the paper with a summary of our results.

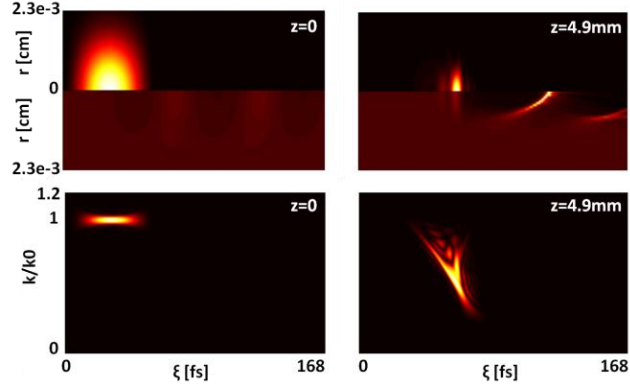

Figure 1 Top: laser intensity and electron density as functions of channel radius r and $\xi$ at propagation distances of z = 0.0, left, and 4.9 mm, right. Bottom: Wigner distributions of the laser pulse as a function of normalized wavenumber and $\xi$ at the same propagation distances.

## II. Plasma Response and Propagation Model

We adopt the ponderomotive guiding center, quasi-static, full wave equation model to describe the evolution of the plasma and laser pulse. The model separates the electric and magnetic fields into fast and slow components. The fast evolution occurs on the time scale of the optical period, and the slow evolution on the time scale of the pulse duration or plasma period. The plasma electrons conserve canonical momentum in responding to the fast components, and respond on the slow time scale to both the slow components of the fields (the plasma wake) and the ponderomotive force associated with the laser pulse. The laser pulse envelope evolves on the slow time scale, responding to changes in the plasma density. On the fast time scale, the electric field is approximated as divergence free. The equations describing the separation of time scales are derived in Ref. 7 for the case of a slowly varying envelope description of the laser pulse. They have been updated in Ref. 22 and are further expanded here to account for large shifts in the spectral content of the laser pulse.

In the model, the transverse component of the laser vector potential for a linearly polarized wave is written as an envelope $\hat{A}$ modulating a plane wave traveling at the speed of light

$$A = \hat{A}(\vec{x}_\perp, \xi, t) e^{-ik_0\xi} + c.c., \tag{1}$$

where $k_0 = \omega_0/c$ is the laser initial central wavenumber. Here a moving window frame is assumed with associated coordinate $\xi = ct - z$. Substituting (1) into Maxwell's Equations for the fast varying field components, assuming the fast plasma current is determined by the ponderomotive guiding center model, and assuming the fast electrostatic field can be neglected ($\nabla \cdot \vec{A} \simeq 0$), the resulting wave equation is

$$\left[\frac{2}{c}\frac{\partial}{\partial t}\left(ik_0 - \frac{\partial}{\partial \xi}\right) - \frac{1}{c^2}\frac{\partial^2}{\partial t^2} + \nabla_\perp^2\right]\hat{A} = \frac{4\pi e^2}{m_e c^2}\left\langle\frac{\bar{n}_e}{\bar{\gamma}}\right\rangle \hat{A}, \tag{2}$$

where $\bar{n}_e$ and $\bar{\gamma}$ are the average electron density and relativistic factor, $e$ is the fundamental unit of charge, $m_e$ is the electron mass, and the angular bracket signifies that an average over an ensemble of particles that stream through the plasma wake generated by the laser pulse is to be taken.

We note that $k_0$ appearing in Eqs. (1) and (2) is a reference wave number, and that the evolution of the vector potential (1) is unchanged by a transformation that shifts the value of $k_0$ and adds a wavenumber shift to the phase of the envelope. This is because the slowly varying plasma density responds only to the magnitude of the envelope, $\hat{A}$. Thus, the validity of Eqs. (1) and (2) does not require that the envelope be slowly varying on the scale of the initial wavelength, $|\partial \hat{A}/\partial \xi| \ll |k_0 \hat{A}|$. However, the multiple time scale approach does require that the pulse envelope evolves slowly as it propagates $|(ik_0 - \partial/\partial \xi)\hat{A}| \gg |\partial \hat{A}/c\partial t|$. The advantage of retaining the second order time derivative and the numerical implementation of Eq. (2) are described further in Appendix A.

## III. Spectral Shifting in Tenuous Plasma

As a laser pulse propagates through plasma, it loses energy by exciting plasma waves[2-4,19,23]. At the same time the laser pulse spectrum is modified. These processes are related to each other through the conservation of wave action. A consequence of the ponderomotive guiding center description of the plasma response is that the plasma current on the right hand side of Eq. (2) is proportional to, and in phase with, the vector potential. The rate of energy transferred to the electromagnetic wave can be calculated as the power the ponderomotive force does on the plasma current[23]. It follows from this that the action,

$$I = \text{Re}\left\{ \int \frac{d^2 x_\perp d\xi}{2\pi c} \hat{A}^*(k_0 + i\frac{\partial}{\partial \xi} + i\frac{\partial}{c\partial t})\hat{A} \right\}, \quad (3)$$

is conserved[22]. We compare this quantity with twice the energy stored in the fast varying electric field,

$$U_L = \int d^2 x_\perp \int \frac{d\xi}{2\pi} \left| (k_0 + i\frac{\partial}{\partial \xi})\hat{A}(\vec{x}_\perp, \xi, t) \right|^2 = \int d^2 x_\perp \int dk k^2 |\bar{A}|^2, \quad (4)$$

representing the sum of the electric and magnetic field energies which are approximately equal. The over bar denotes a Fourier transform with respect to $\xi$ with the transform variable $k$,

$$\bar{A}(\vec{x}_\perp, k, t) = \int d\xi \hat{A}(\vec{x}_\perp, \xi, t) e^{-ik\xi}. \quad (5)$$

The average wavenumber, $<k>$, can be defined and used to monitor the spectral shift. In particular, we define the average wavenumber as

$$<k> = \frac{\int d^2 x_\perp \int dk k^2 |\bar{A}|^2}{\int d^2 x_\perp \int dk k |\bar{A}|^2} \simeq \frac{U_L}{I}. \quad (6)$$

Equation (6) demonstrates that while the action is constant, the average wavenumber must decrease in proportion to the pulse energy. An expression for the rate of energy depletion and accompanying spectral shifting can be derived when plasma waves are weakly excited. For an initial pulse envelope

$$\hat{A}(\vec{x}_\perp, \xi, t=0) = a_0 \sin(\pi\xi/2c\tau) \exp(-r^2/w^2) \quad (7)$$

defined on the interval $0 < \xi < 2c\tau$, where $a_0$ is the peak normalized vector potential of the laser field, $w$ is the spot size, and $\tau$ the full width at half maximum (FWHM). The spectral shifting rate $<k>^{-1} d<k>/dz$ for linear plasma wave excitation is given by

$$\frac{1}{<k>}\frac{d<k>}{dz} = -\frac{1}{2\pi}\frac{1}{c\tau(k_0 w)^4}\left(\frac{P}{P_*}\right)\left[1+\left(\frac{k_p w}{2}\right)^2\right]\left[\frac{\sin(\omega_p \tau)}{1-(\omega_p \tau/\pi)^2}\right]^2, \qquad (8)$$

where $P[GW] \approx 21.5(a_0 w/\lambda)^2$, $P_* = 0.345 GW$, $\omega_p^2 = 4\pi e^2 n_e/m_e$, and $k_p = \omega_p/c$. The dependence of Eq. (8) on physical parameters can be explained as follows. By increasing the pulse power or decreasing the spot size, the work done by the laser pulse on the plasma increases through an enhanced ponderomotive force. For small pulse length, $\omega_p \tau < 1$, the shifting rate is proportional to pulse length as the ponderomotive force is active over an extended duration and performs more work on the plasma. At $\omega_p \tau \sim 2$, the spectral shifting rate is maximized due to the weak resonance associated with matching the pulse length to the plasma period. Increasing the pulse length beyond $\omega_p \tau \sim 2$ provides additional pulse energy but weakens the longitudinal ponderomotive force, diminishing the shifting rate.

While Eq. (8) provides some qualitative insight on the parametric dependence of the shifting rate, its quantitative predictive capability is limited. Figure 2 shows the average wavenumber as a function of distance for a 17TW, 40TW, and 60 TW, laser pulse propagating in a plasma channel with on axis density $4.3 \times 10^{18}$ cm$^{-3}$. The pulse was initialized with a wavelength of 800 nm, temporal FWHM of 30 fs, and a spot size of 25 μm matched to the plasma density's radial parabolic profile[24]. The reduced group velocity accompanying the red-shifting causes the radiation to slide backwards in the moving frame coordinate, $\xi$. Each curve terminates when the radiation reaches the simulation boundary in $\xi$ of 288 fs. The rate of spectral red-shifting increases with power, but the wavenumber does not drop linearly with distance as predicted by Eq. (8). The inset displays a comparison of Eq. (8) with the simulation results for the initial stage of propagation. For the comparison, Eq. (8) has been corrected with the effective relativistic mass increase contributed by the electron quiver energy ($\omega_p \to \omega_p /[1+2a_0^2]^{1/4}$). For lower powers and short propagation distances, the linear theory matches the simulations well. The onset of pulse self-focusing and compression, and electron cavitation result in the deviation between theory and simulation for higher powers and longer propagation distances. Furthermore, as seen in Figs. (1) and (2), the average wavenumber cannot diagnose the spectral range of the pulse. The average wavenumber for the 17 TW pulse in Fig. (2) terminates at $<k>/k_0 \sim 0.5$ corresponding to $\lambda \sim 1.6 \ \mu m$, while in Fig. (1) the spectral range for the same power extends below $k/k_0 \sim 0.2$, $\lambda \sim 4 \ \mu m$. These strongly shifted components of the pulse represent the MIR radiation, wavelengths from 3 to 7 $\mu m$.

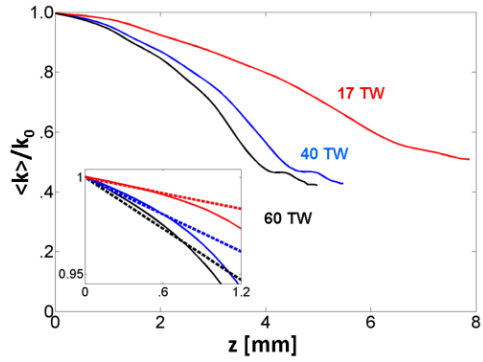

Figure 2: Average wavenumber <k>/k$_0$ as a function of distance z for initial pulse powers of 17, 40 and 60 TW, represented by red, blue and black respectively. The inset shows a comparison with linear theory (dashed lines) for each case.

## IV. Simulation Results

### A. Observation of MIR

We first examine the conversion of spectral energy into the mid-IR for parameters relevant to the current laser-plasma capabilities at the University of Maryland. A laser pulse with an initial wavelength 800 nm, spot size 25 μm, temporal FWHM 30 fs and power 17 TW (normalized vector potential $a_0 = 0.45$), is injected into a plasma channel with an on axis electron density of $4.3 \times 10^{18}$ cm$^{-3}$ and propagates over a total distance of 12.3 mm. The corresponding critical power for self-focusing, $P_{cr} = 17(\omega/\omega_p)^2$ GW, is 7 TW. In Fig. 3, the spectral energy density, the integrand of Eq. (4), is plotted as a function of normalized wavenumber, k/k$_0$, at propagation distances of z = 0, 2.46, 4.94 and 7.9 mm, represented by the color red, grey, dark grey and black respectively. For reference, the pulse energy has dropped to 55% of the initial energy by 7.9 mm. The spectral density is broadened and red shifted as the pulse propagates. At 7.9 mm, the spectral intensity extends to wavelengths longer than 4.7 µm (red arrow).

The blue curve in Fig. 3 displays the spectral density at z = 7.9 mm when the pulse propagates through a uniform plasma (no pre-formed channel). In spite of surpassing the self-focusing condition, $P/P_{cr} = 2.4$, the pulse diffracts in the uniform plasma, limiting the spectral broadening and MIR energy. At 17 TW, the radial confinement provided by pre-formed channel maintains the pulse intensity over an extended distance, allowing additional spectral broadening and increasing the mid-IR energy.

Figure 4 displays the cumulative energy as a function of k/k$_0$ at distances of z = 0, 2.46, 4.94 and 7.9 mm in a plasma channel and at z = 7.9 mm in a uniform plasma demarcated by the colors red, grey, dark grey, black and blue respectively. The cumulative energy, the integral of the spectral energy density up to k, represents the energy contained in wavenumbers less than or equal to k. Wavelengths longer than 4.7 µm (red arrow) account for 15 mJ after 7.9 mm of propagation in the plasma channel (black) with a conversion efficiency from the initial pulse energy of 2.9%. Without a preformed channel (blue), the conversion efficiency is zero for wavelengths longer than 2.5 µm.

Figure 5 is the radially averaged intensity profile of the MIR pulse found by spectrally filtering in the range 0<k<0.2k$_0$ as a function of ξ and z. The MIR pulse appears at a propagation distance near 7.9 mm consistent with Figs. (3) and (4) with a temporal FWHM of 30 fs or two cycles of 4.7 μm light. The MIR pulse gradually lengthens due to group velocity dispersion, and quickly falls back in ξ due to the smaller group velocities associated with the longer wavelengths.

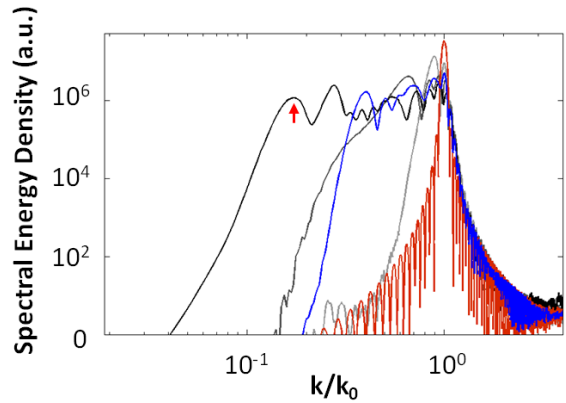

Figure 3 Spectral energy density of a pulse propagating in a plasma channel as a function of $k/k_0$ at distances z=0, 2.5, 4.9 and 7.9 mm, represented by the red, grey, dark grey and black lines respectively; spectral energy density for a pulse propagating through a uniform plasma (no channel) is represented by the blue curve. The red arrow indicates where λ=4.7 μm.

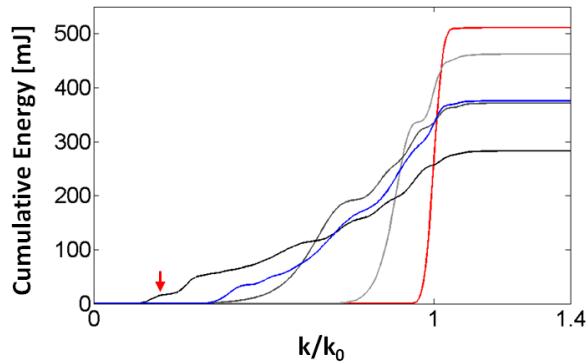

Figure 4 Cumulative energy as a function of $k/k_0$ at distances z=0, 2.5, 4.9 and 7.9 mm, represented by the red, grey, dark grey and black curves respectively; cumulative energy for the no channel case at z=7.9 mm represented by the blue curve . The red arrow indicates there are 15mJ of cumulative energy at λ=4.7 μm.

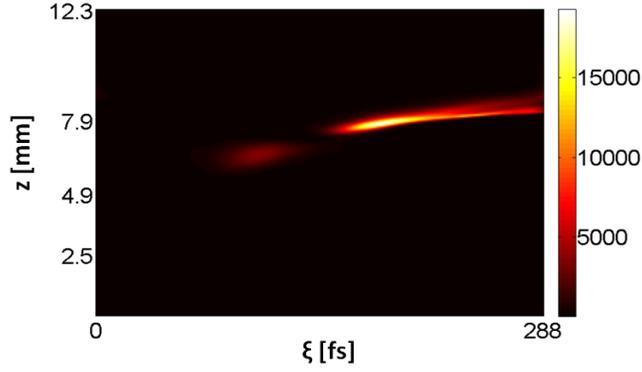

Figure 5 Retrieved profile of the filtered mid-ir pulse ($0<k<0.2k_0$) radially averaged intensity as a function of $(z,\xi)$.

## B. Parametric Dependence of MIR Energy Generation

Maximization of the MIR energy is essential for its use in applications. To this end, we now investigate the parametric dependence of energy conversion to MIR on initial pulse power, pulse duration, and plasma density. Because the spectral shifting rate depends on all of these parameters, all results are plotted at the distance where the average wavenumber shift, Eq. (6), is maximized. All scalings are performed in a plasma channel with matched spot size of 25 μm unless otherwise state. The overall trend is that MIR efficiency increases with the ponderomotive force up to electron cavitation where the efficiency saturates.

The pulse power at fixed pulse length is varied through the laser pulse amplitude while the other parameters remain unchanged from Sec IV. Figure 6 (a) shows the cumulative energy as a function of $k/k_0$ for initial powers of 40 TW, black line, and 17 TW, grey line, with and without a channel, the solid and dashed lines respectively. The larger pulse powers clearly increase the energy content in the MIR range. As discussed above, the pre-formed channel keeps the 17 TW pulse radially confined extending the distance over which the pulse undergoes spectral broadening. At 40 TW, $P/P_{cr} = 5.7$, self-focusing ensures radially confinement of the pulse and the pre-formed channel is no longer required: the cumulative MIR energy resulting from channeled and non-channeled propagation is nearly identical.

Figure 6 (b) displays the conversion efficiency, the cumulative energy divided by the initial laser pulse energy, as a function of initial laser pulse power for the wavelength ranges $\lambda \geq 6$ µm, $\lambda \geq 4$ µm and $\lambda \geq 2$ µm. The efficiency exhibits a threshold behavior with power. The weak power dependence of efficiency above threshold results from relativistic mass increases to the electrons, nonlinear laser pulse evolution, and saturation of the index of refraction gradient through electron cavitation. The electron quiver momentum and consequently the effective mass increase with power. A larger ponderomotive force is required for the same displacement in electron density. As a result, the scaling of electron density gradient with pulse power weakens as power is increased: $\partial_\xi n_e \propto P$ for $a_0 \ll 1$ and $\partial_\xi n_e \propto P^{1/2}$ $a_0 \gg 1$.

For the density considered, the critical power for self-focusing is $P_{cr} = 7\ TW$. The 17 TW and 80 TW pulses undergo stronger nonlinear self-focusing, which increases the ponderomotive force. In addition to transverse compression, spectral broadening of the pulses causes self-steepening or temporal compression. The group velocity decreases as the wavelength increases in a plasma, so that initial red-shifting at the front of the pulse causes pulse energy to coalesce at the back of the pulse. The enhanced amplitude and reduced pulse length both contribute to an increased ponderomotive force. As the pulse becomes radially and temporally localized, the ponderomotive force expels an increasing number of electrons. Eventually the ponderomotive force is large enough to expel all the electrons, cavitation. Further increasing the power causes cavitation earlier in the pulse, increasing the index of refraction gradient, but at the same reduces the temporal region of the pulse that undergoes spectral broadening.

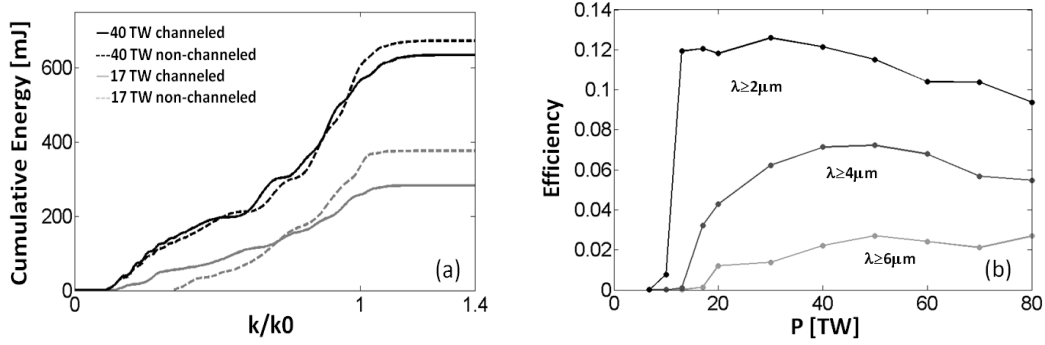

Figure 6 (a) Cumulative energy at largest average wavenumber shift as a function of $k/k_0$ for initial powers P=40 for the channeled (black, solid) and non-channeled (black, dashed) cases and 17TW for the channeled (grey, solid) and non-channeled (grey, dahed) cases. (b) Mid-ir generation efficiency as a function of initial pulse power for $\lambda \geq 6$ μm, $\lambda \geq 4$ μm and $\lambda \geq 2$ μm.

To examine the dependence on pulse length, we fix the pulse energy at 0.5 J. For comparison with the previous case, this can be considered a power scaling at fixed energy. As before the on-axis plasma density is $4.3 \times 10^{18}$ cm$^{-3}$ corresponding to a resonant FWHM $\tau_R = \pi/\omega_p = 27 fs$. Figure 7 shows the conversion efficiency as a function of initial pulse length for the wavelength ranges: $\lambda \geq 6$μm, $\lambda \geq 4$μm and $\lambda \geq 2$μm. The conversion efficiency drops with increasing pulse length. As discussed above, for pulse durations longer than the resonant FWHM, the longitudinal ponderomotive force weakens with increasing pulse length. Additionally, the initial power is decreased as pulse length is increased: $P \sim 1/\tau$. The drop in power weakens both components of the ponderomotive force. A reduced ponderomotive force provides less charge displacement, a smaller index gradient, and less spectral broadening. The dashed lines above and below the $\lambda \geq 2$μm curve illustrate the minor effect of adding a negative and positive chirp respectively. For each pulse duration, the chirp is calculated assuming a 30 fs bandwidth limited pulse.

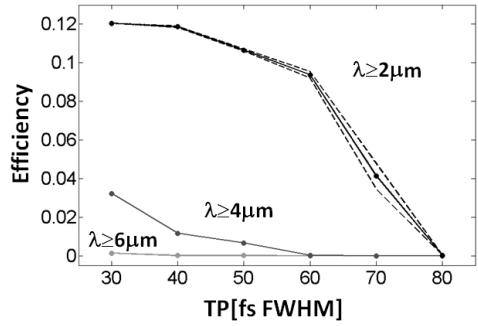

Figure 7: MIR conversion efficiency in the ranges λ ≥ 6μm, λ≥4μm and λ≥2μm as a function of initial temporal FWHM. The upper and lower dashed lines show the effect of negative and positive chirp respectively.

Finally we consider the dependence of MIR generation on the on-axis plasma density. The laser power and temporal FWHM are fixed at 17 TW and 30 fs respectively. The resonant density, defined by $\omega_p = \pi/\tau$, is initially $3.4 \times 10^{18} cm^{-3}$. Figure 8 shows the conversion efficiency as a function of plasma density for the wavelength ranges λ≥6 μm, λ≥4 μm and λ≥2 μm. The critical power varies from 10 TW at $3 \times 10^{18} cm^{-3}$ to 4.6 TW at $6.5 \times 10^{18} cm^{-3}$. The increase in efficiency from $3 \times 10^{18} cm^{-3}$ to $3.4 \times 10^{18} cm^{-3}$ for λ≥2 μm can be attributed to the enhanced plasma wave amplitude at the resonant density. As the density is increased, the drop in critical power provides additional self-focusing and an enhanced ponderomotive force, more charge displacement, and increased spectral broadening observable in the increased efficiency for λ≥4 μm and λ≥6 μm.

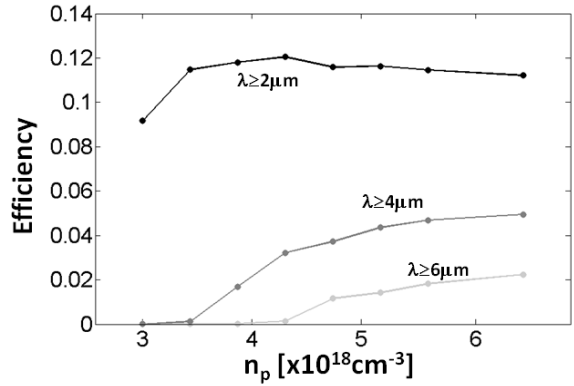

Figure 8: MIR conversion efficiency in the ranges λ≥6 μm, λ≥4 μm and λ≥2 μm as a function of on-axis plasma density.

## V. Conclusions

We have investigated the generation of MIR radiation from spectral broadening of high intensity, ultrashort laser pulses propagating through tenuous plasma. The ponderomotive guiding center, quasi-static, full wave equation model was adopted to simulate the laser pulse and plasma evolution. The conversion of optical energy to the MIR was examined as a function of laser pulse power, pulse length, and plasma density. The predominant trend was for the conversion efficiency to increase with the ponderomotive force up to the point of electron cavitation where the efficiency saturated. The simulations show 1% conversion to MIR energy at wavelengths longer than 6 microns for laser powers greater than 20 TW.

## Acknowledgements

The authors would like to thank P. Sprangle, L. Johnson, T. Rensink, C. Miao, and S.J. Yoon for useful discussions. We would also like to thank support of NSF, ONR and DOE for their support.

## Appendix A: Enveloped Full Wave Equation

The previously used modified paraxial equation (MPE) results from neglecting the second order time derivative in Eq. (2)[22]. Here we retain the second derivative in Eq. (2), which we refer to as the full wave equation (FWE). The advantage of retaining this term is best seen by comparing the linear dispersion relation for the various models. The dispersion relation for a small amplitude plane wave propagating in plasma is given by

$$\omega = \pm\sqrt{|c\vec{k}|^2 + \omega_p^2}, \qquad (A1)$$

where $\omega$ is the frequency, and $\vec{k}$ the wavenumber. For positive $k_z$ ($k_z = \hat{z}\cdot\vec{k}$), the plus and minus signs represent forward and backward propagating modes respectively. The numerical dispersion relation will depend on the manner in which Eq. (2) is converted to a finite difference equation and the grid resolution in $\xi$ and t, $\Delta\xi$ and $\Delta$t. Our typical resolutions are $k_0\Delta\xi = 0.28$, $\omega_p\Delta t = 0.54$. Figure 9 compares the dispersion relation given by Eq. (A1) with the numerical dispersion relation for our implementation of the FWE and the previously used MPE[4,22]. Here we have taken $\omega_p = 1.17\times 10^{14}\ rad/s$, $k_0 = 7.8\times 10^4\ rad/cm$ and $k_\perp = 8\times 10^2\ rad/cm$, corresponding to a plasma density of 4.3x10$^{18}$ cm$^{-3}$ and vacuum focal spot $w_0$ ($k_\perp \sim 2/w_0$) of 25 μm.

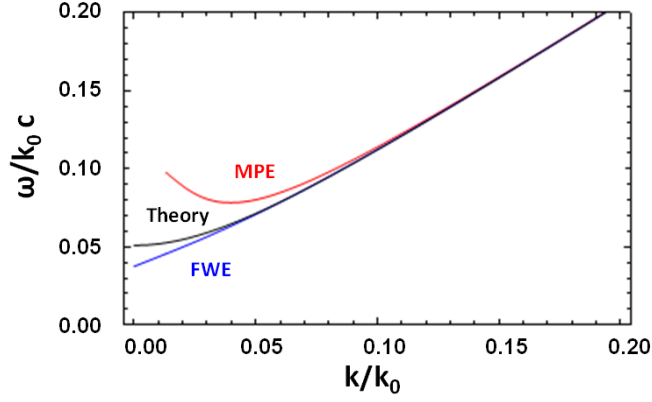

Figure 9 Normalized frequency, ω/k₀c, as a function of normalized wavenumber shift δk/k₀ from the MPE solver (red), FWE solver (blue) and theory (black).

The FWE dispersion agrees well with the theoretical curve for small wavenumber shifts, $|k/k_0| \leq 0.1$, and for larger shifts $|k/k_0| \simeq 0$ the discrepancy is less than 20%. Furthermore, the problematic divergence in the MPE dispersion as $|k/k_0| \to 0$ is eliminated in the FWE. Here, accurate dispersion is particularly important as we are interested in the generation of long wavelength radiation, correspondingly to large shifts $|k/k_0| \simeq 0.1$.

Similar to Ref. 4, each term in Eq. (2) is evaluated at each grid point in $\xi$ and half way between grid points in t

$$\left[\frac{2}{c}\frac{\partial}{\partial t}\left(ik_0 - \frac{\partial}{\partial \xi}\right)\hat{A}\right]_{t+\Delta t/2,\xi} - \frac{1}{c^2}\left[\frac{\partial^2}{\partial t^2}\hat{A}\right]_{t+\Delta t/2,\xi} = \left[\left(k_p^2 + k_\perp^2\right)\hat{A}\right]_{t+\Delta t/2,\xi}. \quad (A2)$$

The differencing for the first and second order time derivatives are evaluated by $[\partial A/\partial t]_{t+\Delta t/2,\xi} = (A_{t+\Delta t,\xi} - A_{t-\Delta t,\xi})/2\Delta t$ and $[\partial^2 A/\partial t^2]_{t+\Delta t/2,\xi} = (A_{t+\Delta t,\xi} - 2A_{t,\xi} + A_{t-\Delta t,\xi})/\Delta t^2$ respectively, and the quantity $\hat{A}_{t+\Delta t/2,\xi}$ on the right of Eq. (A2) is averaged $(\hat{A}_{t+\Delta t,\xi} + \hat{A}_{t-\Delta t,\xi})/2$. This choice of differencing scheme was motivated by its linear stability and dispersion accuracy. To evaluate the derivative with respect to $\xi$, a four point, one sided differencing method is applied[22]

$$\left.\frac{\partial \hat{A}}{\partial \xi}\right|_{\xi} = \frac{1}{\Delta \xi}\left(\frac{5}{3}\hat{A}_{\xi} - \frac{5}{2}\hat{A}_{\xi-\Delta\xi} + \hat{A}_{\xi-2\Delta\xi} - \frac{1}{6}\hat{A}_{\xi-3\Delta\xi}\right). \qquad (A3)$$